\documentclass[twocolumn,showpacs,preprintnumbers,amsmath,amssymb,prb,aps]{revtex4-1}
\usepackage{epsfig}% Include figure files
\usepackage{graphicx}% Include figure files
\usepackage{dcolumn}% Align table columns on decimal point
\usepackage{bm}% bold math
\usepackage{textcomp}
\usepackage{subfig}

\begin{document}

\title{Investigation of Thermoelectric Properties of La$_{0.75}$Ba$_{0.25}$CoO$_{3}$ Compound in High Temperature Region}

\author{Saurabh Singh${{^{1}}}$}
\altaffiliation{Electronic mail: saurabhsingh950@gmail.com}
\author{Devendra Kumar${{^{2}}}$}
\author{Sudhir K. Pandey${{^{1}}}$}
\affiliation{${{^{1}}}$School of Engineering, Indian Institute of Technology Mandi, Kamand - 175005, India}

\affiliation{${{^{2}}}$UGC-DAE Consortium for Scientific Research, University Campus, Khandwa Road, Indore-452001,India}

\date{\today}
\begin{abstract}
In the present work, we have reported the temperature dependent thermopower ($\alpha$) behavior of La$_{0.75}$Ba$_{0.25}$CoO$_{3}$ compound in the temperature range 300-600 K. Using the Heikes formula, the estimated value of $\alpha$ corresponding to high-spin configuration of Co$^{3+}$ and Co$^{4+}$ ions is found to be $\sim$$16$ $\mu$V/K, which is close to the experimental value, $\sim$13 $\mu$V/K, observed at $\sim$600 K. The temperature dependent TE behavior of the compound is studied by combining the WIEN2K and BoltzTrap code. The self consistency field calculations show that the compound have ferromagnetic ground state structure. The electronic structure calculations give half metallic characteristic with a small gap of $\sim$50 meV for down spin channel. The large and positive value for down spin channel is obtained due to the unique band structure shown by this spin channel. The temperature dependent relaxation time for both the spin-channel charge carriers is considered to study the thermopower data in temperature range 300-600 K. An almost linear values of $\tau_{up}$ and a non-linear values of  $\tau_{dn}$ are taken into account for evaluation of $\alpha$. By taking the temperature dependent values of relaxation time for both the spin channels, the calculated values of $\alpha$ using two current model are found to be in good agreement with experimental values in the temperature range 300-600 K. At 300 K, the calculated value of electrical conductivity by using the same value of relaxation time, i.e. 0.1 $\times $10$^{-14}$ seconds for spin-up and 0.66 $\times $10$^{-14}$ seconds for spin-dn channel, is found to be equal to the experimentally reported value. 

\end{abstract}

%\pacs{71.20.-b, 71.15.Mb, 74.25.Fy}

\maketitle

\section{Introduction} 
In 1997, the discovery of good thermoelectric properties in the Na$_{x}$CoO$_{3}$ by Terasaki \textit{et al.} gave a new breakthrough for oxide materials as potential TE candidates.\cite{Terasaki} In the last few decades, the TE properties exhibited by the strongly correlated electron systems have been very interesting topic from physics as well as  industrial and engineering point of view.\cite{Koumoto} The theoretical study of thermopower or Seebeck coefficient ($\alpha$) in strongly correlated system has been carried out by several authors, which is mainly based on the Hubbard model.\cite{Chaikin,Oguri, Doumerc, Mars, koshibae, Palsson, Merino, Jakli} At high temperatures, i.e. normally above 300 K, the temperature independent thermopower has been analysed by considering the Heikes formula or formula given by Chaiken and Beni.\cite{Heikes, Chaikin} At high temperature (T$\rightarrow$$\infty$) the generalized expression of Heikes formula in is given by,\cite{Heikes}
\begin{equation}
  \alpha(T\rightarrow\infty) = -\frac{\textit{k}_{B}}{e} [ln(1-x)/x]
\end{equation}
where, \textit{x} is the concentration of charge carrier, $\textit{k}_{B}$ and \textit{e} are Boltzmann constant and electronic charge, respectively. In Heikes formula, the value of $\alpha$ at high temperature is measure of the entropy per carrier in a system. The above expression of $\alpha$ (i.e. Eq$^{n}$ 1) was unable to adequately explain both the magnitude and temperature dependence in NaCo$_{2}$O$_{4}$. In the strongly correlated systems such as transition-metal oxides, the spin and orbital degrees of freedom of the charge carrier plays an important role in the thermopower. Therefore, consideration of spin and orbital degeneracy are very essential to understand the high temperature thermopower in these compounds.\\
The aspects of spin and orbital degree of freedom were taken into account and Heikes's formula was further extended by Koshibae et al to understand the high temperature thermopower of strongly correlated systems. The modified formula proposed by Koshibae et al is given by,\cite{koshiba}
\begin{equation}
  \alpha(T\rightarrow\infty) = -\frac{k_{B}}{e} ln[(\frac{g_{I}}{g_{II}})(\frac{1-x}{x})]
\end{equation}
where,  g$_{I}$ and g$_{II}$ are the degeneracies and it originates from the spin and orbital degrees of freedom. This expression (Eqn 2) of Koshibae et al. has described the thermopower of many compounds such as NaCo$_{2}$O$_{4}$,\cite{koshibae} Ca$_{3}$Co$_{4}$O$_{9}$,\cite{Nag} cobalt perovskites La$_{1-x}$Sr$_{x}$CoO$_{3}$,\cite{Maignan} layered rhodium oxides,\cite{Pelloquin, Kobayashi} orthochromites,\cite{Marsh, Pal} manganese perovskites Pr$_{0.5}$Ca$_{0.5}$MnO$_{3}$, CaMn$_{0.95}$Nb$_{0.05}$O$_{3}$\cite{Weidenkaff} and double-perovskites Ca(Mn$_{3-y}$Cu$_{y}$)Mn$_{4}$O$_{12}$\cite{Kobayash} and spinels LiMn$_{2}$O$_{4}$,\cite{Sparks} iron based structures SrFeO$_{x}$,\cite{Williams} and RBaCo$_{2}$O$_{5+x}$ (R=Gd, Nd).\cite{Taskin} From the available high temperature thermopower data of these compounds, it is noticed that the theoretical values of $\alpha$ obtained by using the Eq$^{n}$ (2), have good agreement with experimental value of $\alpha$ for a particular temperature only above 300 K, and the temperature at which experimental values and theoretical values are found to be in good agreement varies from compound to compound. However, many of these compounds (CaMn$_{0.95}$Nb$_{0.05}$O$_{3}$, Ca$_{3}$Co$_{4}$O$_{9}$ and etc.)\cite{Weidenkaff, Madre} shows the temperature dependent thermopower behavior above 300 K. Thus, the formulas given by the Heikes, and Koshibae \textit{et al}. were used to understand the thermopower of various compounds at high temperature, but these formula were unable to explain the non-linear and temperature dependent behavior of the many correlated electron systems. Also, for high temperature one can only obtain a fix value of $\alpha$ using Eq$^{n}$ 2. The proper understanding of temperature dependent thermopower behavior of strongly correlated electron systems in high temperature region are still lacking. Therefore, we have tried to understand the high temperature thermopower behavior of strongly correlated system by combining the experimental and theoretical tools, and in the present work, we have taken the barium doped LaCoO$_{3}$ compound as a case study.\\
Transport and magnetic properties of barium doped LaCoO$_{3}$ compound has been studied by several authors.\cite{Patil, Kumar, Devendra, Mandal, Hiroyasu, Kriener} Mandal \textit{et al.} have studied the temperature dependent Seebeck coefficient of this compound up to 335 K. To the best of our knowledge, the high temperature thermopower behavior of La$0.75$Ba$0.25$CoO$3$ compound has not been explored so far. Also, the detailed study of TE properties of this compound using electronic band structure is still lacking in the literature. This give a motivation to explore the high temperature TE properties of this compound.\\
  Here, we report the temperature dependent Seebeck coefficient of La$_{0.75}$Ba$_{0.25}$CoO$_{3}$ compound in the temperature range 300-600 K by using the experimental and density functional theory calculations. At 300 K, the observed value of $\alpha$ is $\sim$7 $\mu$V/K and becomes almost constant (very weak temperature dependent) in the temperature range 300-475 K. Above 475 K, its value increases with temperature and reaches to the value $\sim$13 $\mu$V/K at 600 K. The electronic structure calculations exhibit the ferromagnetic ground structure of the compound with half-metallic character. For the spin-down channel, the calculated energy gap is found to be $\sim$47 meV. Using the BoltzTrap code, temperature dependent electrical conductivities ($\sigma$) and Seebeck coefficient ($\alpha$) have been estimated in the temperature range 300-600 K. Two current model is used to study the temperature dependent behavior of $\sigma$ and $\alpha$. The values of $\sigma$ and $\alpha$ are obtained by taking the suitable values of temperature dependent relaxation time for both up and down channel charge carriers. At 300 K, the total value of $\sigma$ is found to be equal to experimental value reported by Mandal \textit{et al.}\cite{Mandal} A small and almost linear change in relaxation time are taken into account for spin-up channel, whereas consideration of non-linear and large variations in the relaxation time for spin-dn channel gives a very good matching between experimental and theoretical values of $\alpha$ in the temperature range 300-600 K.
  \section{Experimental and Computational details}
  The Polycrystalline samples of La$_{0.75}$Ba$_{0.25}$CoO$_{3}$ were prepared through pyrophoric method.\cite{Pati} The step-by-step synthesis details of barium doped LaCoO$_{3}$ compounds are provided in the earlier work by Devendra \textit{et al}.\cite{Kumar, Devendra} For Seebeck coefficient measurement, the synthesized powder sample was pelletized under the pressure of $\sim$35 Kg/cm$^{2}$. Further, it was sintered at 1100 $^{0}$C for 12 hours. The diameter and thickness of the sample were $\sim$5 mm and $\sim$0.5 mm, respectively. The temperature dependent Seebeck coefficient ($\alpha$) measurement was carried out by using home made setup.\cite{Singh}\\
  We have studied the electronic and thermoelectric properties of compound using the full potential linearized augmented plane-wave (FP-LAPW) method implemented in WIEN2k code in combination with the BoltzTraP code.\cite{Blaha, Madsen} We have used the exchange correlation function within local density approximation (LSDA) of Perdew and Wang.\cite{Perdew}
The self consistency calculations were carried out under virtual crystal approximation. The on-site Coulomb interaction strength, U = 2.75 eV is applied among Co \textit{3d} electrons. A conventional unit cell of LaCoO$_{3}$ contains two formula unit. The barium doping at lanthanum site create holes in the unit cell as per the doping amount. Therefore, to do the 25 percent barium doping at lanthanum site of LaCoO$_{3}$ compound, we have removed 0.50 electrons from the unit cell and assumed that the resultant unit cell are equivalent to the La$_{0.75}$Ba$_{0.25}$CoO$_{3}$. The experimental lattice parameters (a = 5.4549 $\AA$ and c = 13.3194 \AA) of rhombohedral structure described by space group R$\bar{3}$c were used for the calculations.\cite{Devendra} The muffin-tin radii of La, Co and O atoms were fixed to 2.46, 1.97 and 1.69 Bohr, respectively. The matrix size for convergence determined by the \textit{R$_{MT}$K$_{max}$}, and value of this parameter was set equal to 7, where \textit{R$_{MT}$} is the smallest atomic sphere radii and K$_{max}$ is the plane wave cut-off. The self consistency iteration was repeated until the total energy/cell and charge/cell of the system gets converged to less than 10$^{-6}$ Ry and 10$^{-3}$ electronic charge, respectively. For accurate calculation of electronic and transport properties, the k-integration mesh size of 50 $\times$ 50 $\times$ 50 was set. The \textit{lpfac} parameter, which represent the number of k-points per lattice point was kept equal to 5 during the calculation of the transport coefficients.   
\section{Results and Discussion}
Fig. 1 shows temperature dependent Seebeck coefficient data in the temperature range 300-600 K. 
\begin{figure}[htbp]
  \begin{center}
   
   \includegraphics[width=0.40\textwidth]{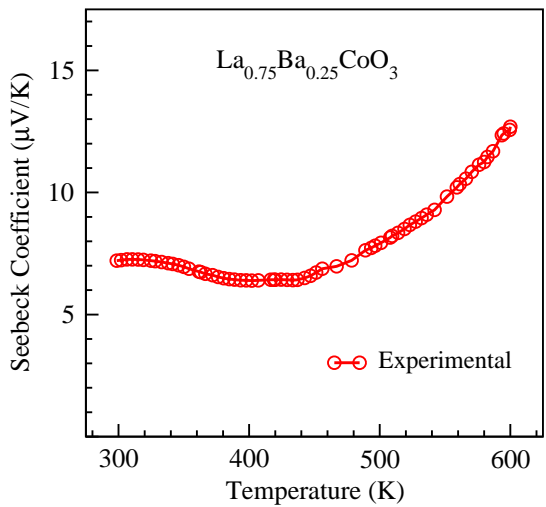}
    \label{}
    \captionsetup{justification=raggedright,
singlelinecheck=false
}
    \caption{(Color online) Temperature dependent Seebeck coefficient of La$_{0.75}$Ba$_{0.25}$CoO$_{3}$ compound.}
    \vspace{0.0cm}
  \end{center}
\end{figure} 
The observed values of $\alpha$ are positive in the entire temperature range under study. At 300 K, the value of $\alpha$ is $\sim$7 $\mu$V/K and it remains almost constant in the temperature range 300-450 K. As the temperature further increases, the values of $\alpha$ increases almost linearly in the temperature range 450-600 K. At 600 K, its value is found to be $\sim$13 $\mu$V/K. In 450-600 K temperature range, a change of $\sim$6 $\mu$V/K is observed in the values of $\alpha$. Normally, a widely known Heikes formula is used to analyze the temperature-independent thermopower.\cite{Chaikin} The general expression of thermopower for cobalt oxides is given by koshibae et al.,\cite{koshibae}
\begin{equation}
  \alpha = \frac{-k_{B}}{|e|} ln(\frac{g_{3}}{g_{4}} \frac{x}{1-x})
\end{equation}
where, \textit{x} is the carrier concentration of Co$^{4+}$ ions; g$_{3}$ and g$_{4}$ are spin and orbital degeneracies associated with Co$^{3+}$ and Co$^{4+}$, respectively. For Co$^{3+}$ (HS) and Co$^{4+}$ (HS), the values of g$_{3}$ and g$_{4}$ comes out to be 15 and 6, respectively. Using the equation (3) and values of g$_{3}$ and g$_{4}$ corresponding to the HS configuration of Co$^{3+}$ and Co$4+$ ions, the value of Seebeck coefficient for La$_{0.75}$Ba$_{0.25}$CoO$_{3}$ compound is found to be $\sim$16 $\mu$V/K. This value of $\alpha$ is close to experimental data observed at $\sim$600 K temperature. At this point it is important to note that the use of Heikes formula is theoretically valid only at high T, in the limit of infinite temperature, i.e. when all the relevant energies become negligible with respect to \textit{k}T. Also, use of expression gives a constant value of $\alpha$ for high temperature region. This shows that Heikes formula is unable to explain the temperature dependent behavior of $\alpha$ of the La$_{0.75}$Ba$_{0.25}$CoO$_{3}$ compound in 300-600 K temperature range. Therefore, to understand the temperature dependent transport properties of the La$_{0.75}$Ba$_{0.25}$CoO$_{3}$ compound, we have further performed the electronic structure calculations.\\
In order to know the ground state of the system, the self-consistent field calculations on the compound corresponding to non-magnetic and ferromagnetic phase were carried out under virtual crystal approximations. The value of total converged energy corresponds to the ferromagnetic (FM) solution was found to be $\sim$125 meV/f.u. lower than that of non-magnetic solution. This clearly suggests that compound exhibit ferromagnetic ground state. Therefore, to understand the transport properties, we have carried out the total (TDOS) and partial density of states (PDOS) of FM phase. In FM phase, the value of total magnetic moment per formula unit for this compound was found to be $\sim$1.75 $\mu$$_{B}$, where the contributions from La, Co, O atoms and interstitial region are 0.0067, 1.7057, -0.0035 and 0.0483 $\mu$$_{B}$, respectively. This calculated magnetic moment per formula unit is $\sim$0.25 $\mu$$_{B}$ larger than that of experimentally reported value ($\sim$1.5 $\mu$$_{B}$) of magnetic moment by Voronin \textit{et al}.\cite{Voronin} The value of magnetic moment reported by Voroin et al. is obtained by the neutron powder diffraction data at 300 K, where effect of thermal agitation is also responsible to lower the value of effective magnetic moment. Therefore, in comparison to experimental value of magnetic moment, a larger value of effective moment is expected from the ground state calculation.\\
\begin{figure}[htbp]
  \begin{center}
   \includegraphics[width=0.40\textwidth, totalheight=0.45\textheight]{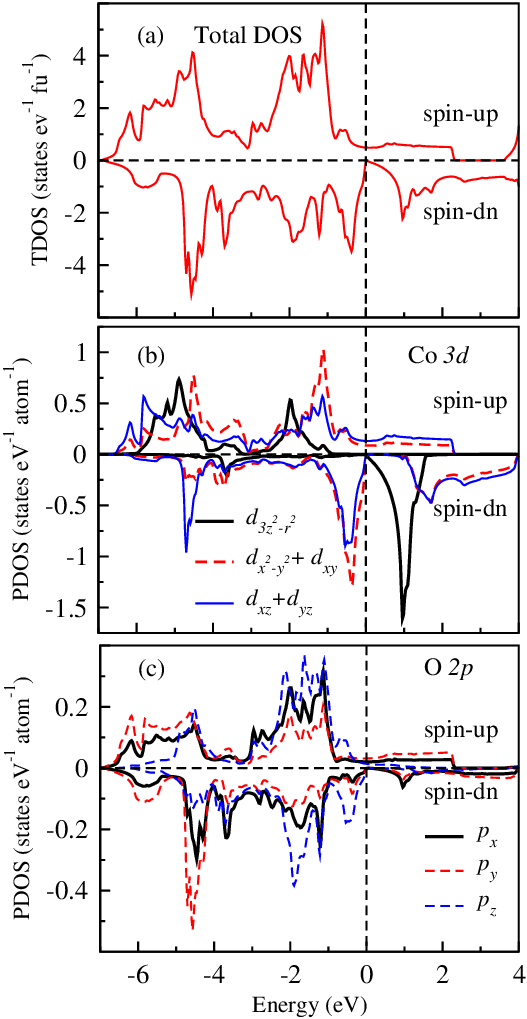}
    \label{}
    \captionsetup{justification=raggedright,
singlelinecheck=false
}
    \caption{(Color online) Total and partial density of states plots for La$_{0.75}$Ba$_{0.25}$CoO$_{3}$. Shown are (a) the TDOS plot, (b) PDOS of Co atom (\textit{3d} orbitals), (c) PDOS of O atom (\textit{2p} orbitals).}
    \vspace{0.0cm}
  \end{center}
\end{figure}
The density of states plots for FM phase are shown in Fig. 2. The dashed line corresponds to 0 eV represents the Fermi level E$_{F}$. From TDOS plot shown in Fig. 2a, it is clearly observed that spin-up channel have $\sim$0.5 states/eV/f.u., whereas for spin-dn channel there is negligibly small DOS at E$_{F}$. Therefore, the character of spin-up and spin-dn channels are metallic and semiconducting, respectively. This suggest that La$_{0.75}$Ba$_{0.25}$CoO$_{3}$ compound exhibit half-metallic behavior. To see the contributions from different atomic states around Fermi level, the PDOS plots for Co \textit{3d} and O \textit{2p} orbitals are shown in Fig. 2b and 2c, respectively. For spin-up channel, the contributions in DOS of Co atom are mainly from the \textit{ d$_{x^{2}-y^{2}}$}+\textit{d$_{xy}$} and \textit{d$_{xz}$}+\textit{d$_{yz}$} orbitals. The values of DOS at E$_{F}$ are $\sim$0.13 and $\sim$0.08 states/eV/atom for \textit{d$_{xz}$}+\textit{d$_{yz}$} and \textit{ d$_{x^{2}-y^{2}}$}+\textit{d$_{xy}$} orbitals, respectively. Thus, due to metallic nature of spin-up channel the electrons in the \textit{ d$_{x^{2}-y^{2}}$}+\textit{d$_{xy}$} and \textit{d$_{xz}$}+\textit{d$_{yz}$} orbitals are expected to give large electrical conductivity, whereas a small and negative Seebeck coefficient is expected from the spin-up channel. For both the channels, at E$_{F}$ the contributions in the DOS of Co atom from d$_{3z^{2}-r^{2}}$ orbitals are negligible. In the PDOS plot of Co atom, the spin-dn channel have two peak in the valence band (VB), one at the edge and another at $\sim$-5 eV of the VB, where as one peak at $\sim$1 eV is present in the CB. The thermally excited electrons from \textit{ d$_{x^{2}-y^{2}}$}+\textit{d$_{xy}$} and \textit{d$_{xz}$}+\textit{d$_{yz}$} orbitals participates in the transport properties. Due to semiconducting nature, large Seebeck coefficient and small electrical conductivity are expected from the spin-dn channel. As the temperature increases, the electrical conductivity of the spin-dn channel also increase due to increase in the population of thermally excited electrons in CB. It is evident from the PDOS plot of O \textit{2p}, the contributions in the DOS at E$_{F}$ from \textit{p$_{x}$}, \textit{p$_{y}$} and \textit{p$_{z}$} orbitals are very small from spin-up channel, and the gap about Fermi level is $\sim$0.7 eV for spin-dn channel. Therefore, very small contribution in the transport properties will be from the O \textit{2p} orbitals. Thus in La$_{0.75}$Ba$_{0.25}$CoO$_{3}$, both channel will contribute in the transport properties. For spin-up and spin-dn channel, the contributions in DOS from different orbitals are different, so different transport behavior is also expected from both the channels. Therefore, the study of transport properties from up and down spin channel is essential, and further estimation of total Seebeck coefficient and electrical conductivity are required to make comparison with experimentally observed value. 
 The spin polarized dispersion curves along the high symmetry directions ($\Gamma$-T-L-$\Gamma$-FB-T) of spin-up and spin-dn channel are shown in Fig. 3a and 3b, respectively. It is also evident from the Fig. 3 that this compound has half-metallic ground state. From Fig. 3a, it is clearly observed that dispersing bands (Band \textbf{\textit{1}} and \textbf{\textit{2}}) crosses the Fermi level E$_{F}$ at 9 different k-points, and the electrons in these bands are responsible for the transport properties of this compound. There will be negligibly small contributions to the transport properties from those bands which do not crosses the Fermi level. The energy of electrons lying in bands \textit{\textbf{1}} and \textit{\textbf{2}} are more than $\sim$1 eV, which is corresponding to the temperature of $\sim$12000 K. Therefore, the electrons in these bands will not affect the transport properties of the compound in the temperature range under study. For spin-dn channel, a clear cut energy gap about Fermi level is observed between band \textit{\textbf{1}} and \textit{\textbf{2}}. The conduction band minimum lies at $\Gamma$ point, whereas maximum of VB lies at a k-point in between $\Gamma$ and FB. This shows that, spin-dn channel have an indirect band gap. Using LSDA exchange correlation functional with U value equal to 2.75 eV, the calculated energy gap for spin-dn channel is found to be $\sim$50 meV, which is $\sim$100 times smaller than that of estimated energy gap ($\sim$0.5 eV) of LaCoO$_{3}$ compound with same U value.\cite{SSingh} At $\Gamma$ point in the VB, the bands \textit{\textbf{2}} and \textit{\textbf{3}} are doubly degenerate and this degeneracy get lifted along the $\Gamma$-FB and $\Gamma$-L directions. 

  \begin{figure}[htbp]
    \captionsetup[subfigure]{labelformat=empty}
\begin{center}

\subfloat[]{
        
        \includegraphics[clip,trim=0.cm 0cm 0.0cm 0.0cm, width=0.40\textwidth]{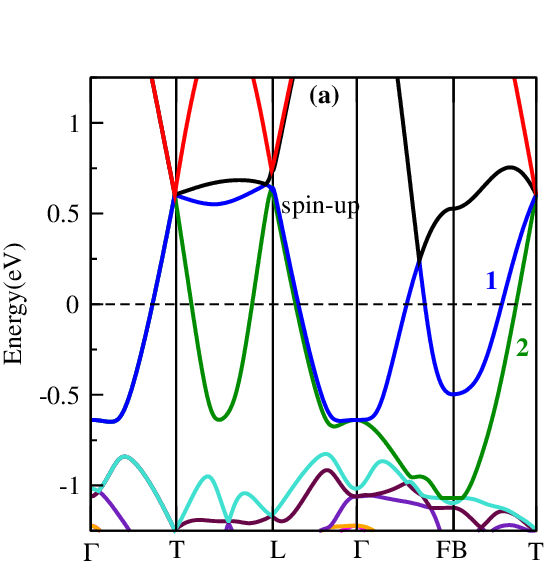} } 
\vspace{-0.85cm}
\subfloat[]{
       
        \includegraphics[clip,trim=0cm 0.0cm 0cm 0.0cm, width=0.40\textwidth]{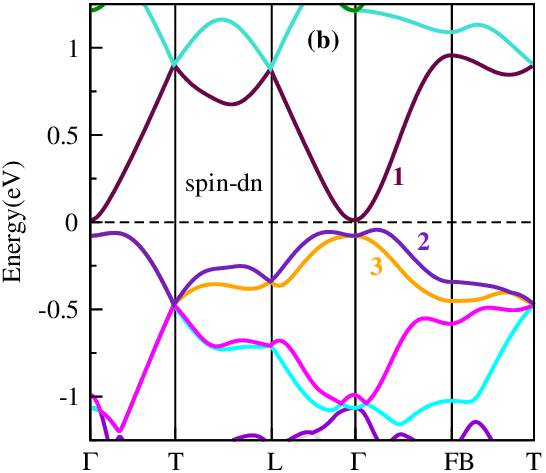} } 
\captionsetup{justification=raggedright,
singlelinecheck=false
}
\caption{(Color online) Electronic band structure of La$_{0.75}$Ba$_{0.25}$CoO$_{3}$ compound, shown (a) spin-up channel (top) and (b) spin-down channel (bottom).}
\label{}
  \vspace{-0.2cm}
\end{center}
\end{figure}
The temperature dependent Seebeck coefficient ($\alpha$) and electrical conductivity per unit relaxation time ($\sigma$/$\tau$) for both spin channels are shown in the Fig. 4. From the Fig. 4(a) and (b), it is clearly observed that the estimated values of $\alpha$ in the temperature range 300-600 K are negative and positive for spin-up and spin-dn channel, respectively. The negative values of $\alpha$ for spin-up channel suggests the presence of \textit{n-type} charge carriers, which is also expected from the metallic character of spin-up channel observed in the band structure plots. For spin-up channel, the value of $\alpha$ at 300 K is $\sim$ -0.4 $\mu$V/K and it decreases almost linearly to the value $\sim$ -1 $\mu$V/K at 600 K. At 300 K, the value of $\alpha$ for spin-dn channel is $\sim$211 $\mu$V/K, which is $\sim$525 times larger than that of spin-up channel. This large and positive value of $\alpha$ is found due to unique band structure shown by down spin channel. It is clearly observed from the Fig. 3b that, an almost equal energy gap of $\sim$50 meV are present in between the bottom of CB (band \textbf{\textit{1}}) and top of the VB (band \textbf{\textit{2}}) at two different points along $\Gamma$ to FB and $\Gamma$ to L directions, and this gap is corresponding to the temperature of $\sim$600 K. At $\Gamma$ point, the energy band gap between top of the band \textit{\textbf{3}} (in VB) and bottom of the band \textit{\textbf{1}} (in CB) is $\sim$77 meV, which is corresponding to the temperature of $\sim$900 K. In the VB, the top of band \textit{\textbf{3}} (at $\Gamma$ point) is $\sim$27 meV (corresponds to $\sim$300 K) lower than the energy corresponding to two peak of Band \textit{\textbf{2}} along $\Gamma$ to FB and $\Gamma$ to L directions. Therefore, a fraction of electrons will be also thermally excited from the band 3, and will do the contributions in the transport properties of the compound in the temperature range 300-600 K. At 300 K, the large number of electrons are thermally excited from the top of band \textit{\textbf{2}} and gets occupied by the available states in C.B. (band \textit{\textbf{1}}). In the transfer process of thermally excited electrons from Vb to CB, the holes are created in the VB. The created holes in the band \textit{\textbf{3}} (at $\Gamma$ point) and band \textit{\textbf{2}} (at peak points along $\Gamma$ to FB and $\Gamma$ to L directions) acts as a positive charge carriers and do the positive contributions in the values of $\alpha$ of the compound. At $\Gamma$ point, the cusp like shape of band \textit{\textbf{2}} suggest that the created holes at $\Gamma$ point act as electron-like and do the negative contributions in the values of $\alpha$. At this point it is important to notice that, at $\Gamma$ point the band \textit{\textbf{3}} is more dispersive in comparison to the band \textit{\textbf{1}} and \textit{\textbf{2}}. Therefore, due to especial band structure shown by spin-dn channel, large and positive values of $\alpha$ are expected from the down-spin channel in the temperature range under study. 

\begin{figure}[htbp]
  \begin{center}
   \includegraphics[width=0.45\textwidth]{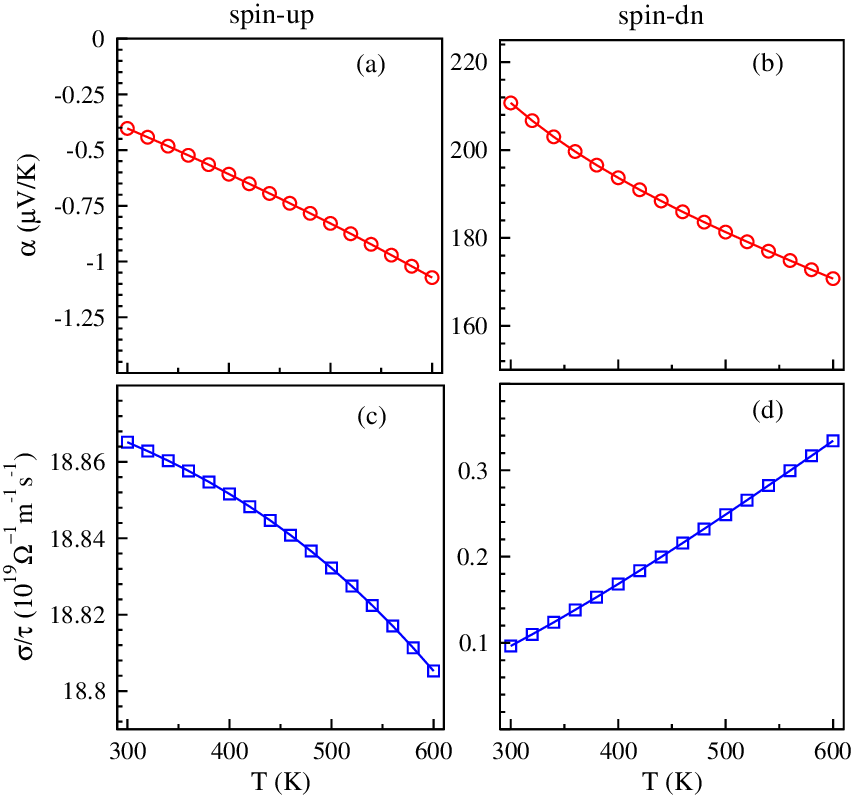}
    \label{}
    \captionsetup{justification=raggedright,
singlelinecheck=false
}
    \caption{(Color online) Variation of transport coefficients with temperature. (a and b) Seebeck coefficient with temperature, (c and d) electrical conductivity with temperature.}
    \vspace{0.0cm}
  \end{center}
\end{figure}
Fig. 4c and 4d shows the variation of $\sigma$/$\tau$ with temperature for spin-up and spin-dn channels, respectively. For the spin-up channel the value of $\sigma$/$\tau$ at 300 K is $\sim$18.90 $\times$ 10$^{19}$ $\Omega$$^{-1}$m$^{-1}$s$^{-1}$. From Fig. 4c, It is clearly observed that the value of $\sigma$/$\tau$ continuously decreases with increase in temperature, which is a typical metal like behavior. At 600 K, it reaches to the value $\sim$18.80 $\times$ 10$^{19}$ $\Omega$$^{-1}$m$^{-1}$s$^{-1}$. The temperature dependent behavior of $\sigma$/$\tau$ for spin-dn channel is shown in Fig. 4d. At 300 K, $\sigma$/$\tau$ is equal to $\sim$0.1 $\times$ 10$^{19}$ $\Omega$$^{-1}$m$^{-1}$s$^{-1}$ and increases almost linearly in the temperature range 300-600 K. At 600 K, its value is $\sim$0.33 $\times$ 10$^{19}$ $\Omega$$^{-1}$m$^{-1}$s$^{-1}$. The temperature dependent behavior of $\sigma$/$\tau$ suggests that this compound have semiconductor-like behavior for down-spin channel. At 300 K, the estimated value of $\sigma$/$\tau$ for spin-up channel is $\sim$190 times larger than that of spin-dn channel. Such large electrical conductivity observed for spin-up channel is due to the availability of large number of free electrons in spin-up channel. The total electrical conductivity ($\sigma$) of the compound is simply sum of the electrical conductivities of spin-up and spin-dn channels. Therefore, the contributions in total electrical conductivity from spin-dn channel is very small.
  \begin{figure}[htbp]
  \vspace{0.3cm}
  \begin{center}
   \includegraphics[width=0.40\textwidth]{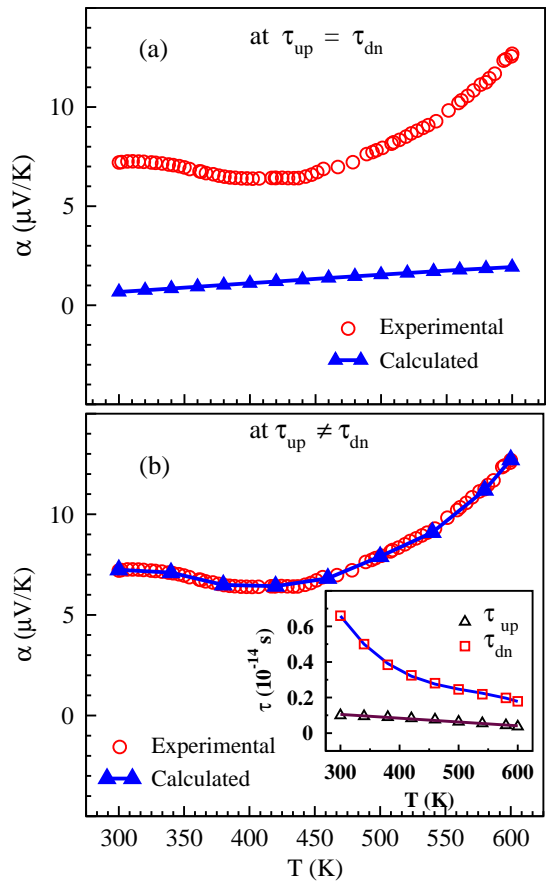}
    \label{}
    \captionsetup{justification=raggedright,
singlelinecheck=false
}
    \caption{(Color online) Variation of total Seebeck coefficient ($\alpha$) with temperature. Comparison of Experimental and calculated $\alpha$, shown (a) for $\tau$$_{up}$ = $\tau$$_{dn}$  and (b) for $\tau$$_{up}$ $\neq$ $\tau$$_{dn}$.}
    \vspace{-0.1cm}
  \end{center}
\end{figure}
  The values of total Seebeck coefficient at different temperatures are obtained by using the two-current model. The temperature dependent variation of experimental and calculated value of $\alpha$ is presented in the Fig. 5. As shown in Fig. 5a, we obtained almost linear temperature dependent bahavior of $\alpha$, by considering the equal relaxation time ($\tau$$_{up}$ = $\tau$$_{dn}$) for both the spin channel. The difference between experimental and calculated values of $\alpha$ are $\sim$7 and $\sim$11$\mu$V/K at 300 K and 600 K, respectively. It is evident from Fig. 5a that neither magnitude nor temperature dependent behavior of $\alpha$ are matching with experimentally observed data in the temperature range 300-600 K. For both metals and semiconductors, the values of relaxation time are normally temperature dependent and its values are typically in the range of 10$^{-14}$ to 10$^{-15}$ seconds.\cite{Ashcroft} Therefore, for understanding the experimentally observed data of $\alpha$, we have further considered the temperature dependent relaxation time for both the spin channel.
  In order to calculate total $\alpha$ using temperature dependent relaxation time, the expression of $\alpha$ given by two current model can be written as,\cite{Xiang, Botana}
  \begin{equation}
    \alpha = [\frac{\alpha \uparrow (\sigma\uparrow/\tau_{up}(T)) + \alpha \downarrow (\sigma\downarrow/\tau_{dn}(T))}{\sigma\uparrow/\tau_{up}(T)+\sigma\downarrow/\tau_{dn}(T)}] 
  \end{equation}
  where, $\tau_{up}(T)$ and  $\tau_{dn}(T)$ are the used value of relaxation time at a given temperature T.\\
   At different temperatures, we have adopted the suitable values of relaxation time for both the channels such that calculated values of $\alpha$ match very closely with experimental values in the temperature range 300-600 K. The temperature dependent behavior of relaxation time for spin-up ($\tau$$_{up}$) and spin-dn ($\tau$$_{dn}$) channels are shown in the inset of Fig. 5b. For spin-up channel, an almost linear behavior in the $\tau$$_{up}$ is noticed, whereas for the spin-dn charge carriers the variations in the $\tau$$_{dn}$ is non-linear. The values of $\tau$$_{dn}$ are larger than that of $\tau$$_{up}$ in the temperature range 300-600 K. The temperature dependent variation in $\tau$$_{dn}$ is very large in comparison to $\tau$$_{up}$ and the value of $\tau$$_{dn}$ approaches to the value of $\tau$$_{up}$ at 600 K, which can be due to the semiconductor to metal transition of spin-dn channel at $\sim$600 K. For the spin-up channel, the used values of $\tau$$_{up}$ at 300 K and 600 K are 0.1 $\times$10$^{-14}$ and 0.037 $\times$10$^{-14}$ seconds, respectively; whereas for the spin-dn channel the values of $\tau$$_{dn}$ at 300 K and 600 K are 0.66 $\times$10$^{-14}$ and 0.18 $\times$10$^{-14}$ seconds, respectively. Using these values of $\tau$$_{up}$ and $\tau$$_{dn}$, we obtained the values of $\alpha$ equal to $\sim$7.24 and $\sim$12.71 $\mu$V/K at 300 and 600 K, respectively. The estimated values of $\alpha$  using these values of relaxation times are very closed to the experimentally observed data.\\ 
    In order to know the details of the temperature dependent variations of relaxation time of both spin-up and spin-dn charge carriers, the $\tau$$_{up}$ and $\tau$$_{dn}$ curves are fitted using the polynomial equations. The best possible fit for $\tau$$_{up}$ and $\tau$$_{dn}$ curves are obtained by using the linear equation (\textit{A$_{0}$} + \textit{A$_{1}$}T) and cubic equation (\textit{B$_{0}$} + \textit{B$_{1}$}T + \textit{B$_{2}$}T$^{2}$ + \textit{B$_{3}$}T$^{3}$), respectively. The values of coefficients obtained from the linear fit are A$_{0}$($\sim$0.17), A$_{1}$($\sim$ -2 $\times$10$^{-4}$) and from cubic fit are B$_{0}$($\sim$4.5, B$_{1}$($\sim$-2.3 $\times$10$^{-2}$), B$_{2}$($\sim$4.4 $\times$10$^{-5}$) and B$_{3}$($\sim$ -2.8 $\times$10$^{-8}$). To see the reliability of the adopted values of relaxation time for both the spin channel, we have also calculated the total electrical conductivity. The electrical conductivity of the half-metallic system is normally sum of the electrical conductivities of both spin channel. Thus, the resultant electrical conductivity of the material can be expressed as,\cite{Dorleijn}
\begin{equation}
  \sigma = \sigma\uparrow + \sigma\downarrow
\end{equation}
where, $\sigma\uparrow$ and $\sigma\downarrow$ are the electrical conductivities of the spin-up and spin-dn channels, respectively. For 300 K, the value of $\sigma$ equal to $\sim$1.95 $\times$ 10$^{5}$ $\Omega$$^{-1}$m$^{-1}$ is obtained by considering the $\tau$ values as 0.1 $\times $10$^{-14}$ and 0.66 $\times $10$^{-14}$ seconds for spin-up and spin-dn channels, respectively. This calculated value of $\sigma$ is found to be in good agreement with the experimentally observed value ($\sim$2 $\times$ 10$^{5}$ $\Omega$$^{-1}$m$^{-1}$).\cite{Mandal} To the best of our knowledge, the experimental values of relaxation time and electrical conductivities in high-temperature region are not available for this compound. Therefore, the comparison of adopted relaxation time and calculated values of electrical conductivities by using these relaxation time are not possible in the present study. It will be interesting to see whether the adopted values of relaxation time here are same to the experimental data of relaxation time and electrical conductivities or not. Thus, the further study of high temperature electrical conductivities and relaxation times are desirable.\\
\section{Conclusions} 
In conclusion, we have investigated the temperature dependent thermopower of La$_{0.75}$Ba$_{0.25}$CoO$_{3}$ compound in the temperature range 300-600 K. The value of $\alpha$ at 300 K is found to be $\sim$7 $\mu$V/K and remains almost constant up to 450 K. Above 450 K, the value of $\alpha$ increases with increase in temperature and reached to $\sim$13 $\mu$V/K at 600 K. For the high-spin configuration of Co$^{+3}$ and Co$^{4+}$ ions, the value of $\alpha$ estimated by Heikes formula comes to be $\sim$16 $\mu$V/K, which is close to the experimentally observed value at 600 K. The temperature dependent behavior of $\alpha$ is studied by using the electronic structure calculations. The self consistency field calculation shows that La$_{0.75}$Ba$_{0.25}$CoO$_{3}$ compound have ferromagnetic ground state structure. The large and positive value of $\alpha$ obtained for the down spin channel is due to unique dispersion curve shown by the spin-dn channel. A good match between experimental and calculated values of $\alpha$ in 300-600 k is obtained by taking into account of the temperature dependent values of relaxation time for both the channels. At 300 K, the values of relaxation time used for up and down channels, 0.1 $\times $10$^{-14}$ and 0.66 $\times $10$^{-14}$ seconds also gives a very good match between calculated and experimentally reported value of electrical conductivity. Thus, the consideration of temperature dependent relaxation time for both the spin-channel plays an important role in the understanding of TE properties of this compound.

\end{document}